\documentclass[11pt,a4paper]{article}
\usepackage{amsfonts}
\usepackage{amsmath,amssymb}
\usepackage{epsfig,graphicx}

\input{tcilatex}
\begin{document}

\begin{center}
\bigskip

{\Huge Emergent photons and gravitons}

{\Huge \bigskip }

\bigskip

\bigskip

\textbf{J.L.~Chkareuli, J. Jejelava and Z. Kepuladze}

\bigskip

\textit{Center for Elementary Particle Physics, Ilia State University, 0162
Tbilisi, Georgia\ \vspace{0pt}\\[0pt]
}

\textit{and} \textit{E. Andronikashvili} \textit{Institute of Physics, 0177
Tbilisi, Georgia\ }

\bigskip \bigskip \bigskip \bigskip \bigskip

\textbf{Abstract}

\bigskip
\end{center}

Now, it is already not a big surprise that due to the spontaneous
Lorentz invariance violation (SLIV) there may emerge massless vector
and tensor Goldstone modes identified particularly with photon and
graviton. Point is, however, that this mechanism is usually
considered separately for photon and graviton, though in reality
they appear in fact together. In this connection, we recently
develop the common emergent electrogravity model which would like to
present here. This model incorporates the ordinary QED and tensor
field gravity mimicking linearized general relativity. The SLIV is
induced by length-fixing constraints put on the vector and tensor
fields, $A_{\mu }^{2}=\pm M_{A}^{2}$ and $H_{\mu \nu }^{2}=\pm
M_{H}^{2}$ ($M_{A}$ and $M_{H}$ are the proposed symmetry breaking
scales) which possess the much higher symmetry then the model
Lagrangian itself. As a result, the twelve Goldstone modes are
produced in total and they are collected into the vector and tensor
field multiplets. While photon is always the true vector Goldstone
boson, graviton contain pseudo-Goldstone modes as well. In terms of
the appearing zero modes, theory becomes essentially nonlinear and
contains many Lorentz and CPT violating interaction. However, as
argued, they do not contribute in processes which might lead to the
physical Lorentz violation. Nonetheless, how the emergent
electrogravity theory could be observationally differed from
conventional QED and GR theories is also briefly discussed.

\bigskip \bigskip

\bigskip

\bigskip \bigskip

\bigskip

\bigskip \bigskip

\bigskip

\bigskip \bigskip

\bigskip

\bigskip

{\tiny Invited talk at the 21st International Workshop "What Comes Beyond
the Standard Model?" (23-30 June 2018, Bled, Slovenia)}

\thispagestyle{empty}\newpage

\section{Introduction}

\label{s:intro}While Lorentz symmetry looks physically as an absolutely
exact spacetime symmetry, the spontaneous Lorentz invariance violation
(SLIV) suggests a beautiful scenario where massless vectors and/or tensor
fields emerge as the corresponding zero modes which may be identified with
photons, gravitons and other gauge fields \cite{bjorken,ohan,eg}. Though
they appear through \ condensation of the pure gauge degrees of freedom in
the starting theory their masslessness are provided by their Nambu-Goldstone
nature \cite{NJL,cfn,jb,kraus,jen,bluhm,kan,kos,car} rather than a
conventional gauge invariance.

\subsection{Emergent vector fields theory}

In order to violate Lorentz invariance one necessarily needs field(s) being
sensitive to the spacetime transformations, as vector or tensor fields are.
They can evolve vacuum expectation value which fixes direction of the
violation in the spacetime and create the corresponding condensate.
Therefore, if there is an interaction with this condensate one could expect
Lorentz violation to be manifested physically. If we want to arrange
spontaneous Lorentz violation by the vector field, we could start, as usual,
with the potential terms in the Lagrangian%
\begin{eqnarray}
L &=&-\frac{1}{4}F_{\mu \nu }^{2}-V\text{ ; \ \ \ \ \ }V=\lambda \left(
A_{\mu }^{2}-n_{\mu }^{2}M_{A}^{2}\right) ^{2}  \label{1} \\
F_{\mu \nu }^{2} &=&F_{\mu \nu }F^{\mu \nu }\text{ ; \ \ \ \ }A_{\mu
}^{2}=A_{\mu }A^{\mu }\text{ ; \ \ }n_{\mu }^{2}=n_{\mu }n^{\mu }  \notag
\end{eqnarray}%
where $n_{\mu }$ is an unit constant vector specifying character of Lorentz
violation. If $n_{\mu }$ is time-like vector, we have time-like violation
breaking $SO(1,3)$ to $SO(3)$. If $n_{\mu }$ is space-like vector, we have
space-like violation breaking $SO(1,3)$ to $SO(1,2)$. \

We started with gauge invariant kinetic term, but since potential violates
gauge invariance anyway, we could have started with general kinetic terms%
\begin{equation}
L_{k}=a\left( \partial _{\alpha }A_{\beta }\right) ^{2}+b\left( \partial
_{\alpha }A^{\alpha }\right) ^{2}
\end{equation}%
but problem arising here is a propagating ghost mode, which we get ride off
with the gauge invariant form of kinetic terms.

Such a system of vector field with potential, generally appears not stable,
its energy is not bound from below unless phase space is restricted with
condition
\begin{equation}
A_{\mu }^{2}-n_{\mu }^{2}M_{A}^{2}=0  \label{c1}
\end{equation}%
While this condition may appear out of the blue, it is actually motivated by
the conserved current of (\ref{1})%
\begin{equation}
J_{\mu }=A_{\mu }(A_{\alpha }^{2}-n_{\alpha }^{2}M_{A}^{2})
\end{equation}%
and if in the initial condition the conserved charge of this current is set
to zero , which means (\ref{c1}) is always zero, no propagating ghosts,
Hamiltonian is positively defined and Coulomb law stays the same \cite{kkk}.
So, basically we arrived to the point where we accept to take $\lambda $ in (%
\ref{1}) to infinity as a Lagrange multiplier and get conventional vector
field kinetics with the addition of (\ref{c1}) condition. This condition
still is a cause for spontaneous Lorentz invariance violation, but in
contrast now Higgs mode is set to zero. This was Nambu's original idea \cite%
{nambu}. It is easy to see, if we write expansion of the vector field $A$
into Goldstone and Higgs modes in the exponential manner, which is%
\begin{equation}
A_{\mu }=(M_{A}+h)n_{\nu }\exp J_{\mu }^{\nu }  \label{2}
\end{equation}%
where $h$ is Higgs mode and Goldstone modes $a_{\mu }$ are sitting in $%
J_{\mu }^{\nu }$ (generators for Lorentz transformation) and $a_{\mu
}=M_{A}n_{\nu }J_{\mu }^{\nu }$, \ $a_{\mu }n^{\mu }=M_{A}n^{\mu }n_{\nu
}J_{\mu }^{\nu }=0$. \ So, \
\begin{equation}
A_{\mu }^{2}=(M_{A}+h)^{2}n_{\nu }^{2}=n_{\alpha }^{2}M_{A}^{2}\text{ \ }%
\Longrightarrow \text{ \ }h=0
\end{equation}

Expansion (\ref{2}) is nonlinear with respect to vector Goldstone modes, but
$\frac{a_{\mu }}{M_{A}}$ is a small parameter and we can expand exponent in
the power series and in the second approximation get
\begin{equation}
A_{\mu }=\left( M_{A}-\frac{n_{\alpha }^{2}a_{\alpha }^{2}}{2M_{A}}\right)
n_{\mu }+a_{\mu }  \label{3.0}
\end{equation}

It is clear now that we get nonlinear Lagrangian for vector Goldstone modes,
which in the first approximation is
\begin{equation}
L(A)\rightarrow L(a)=-\frac{1}{4}f_{\mu \nu }f^{\mu \nu }-\frac{1}{2}\delta
(n_{\alpha }a^{\alpha })^{2}-\frac{1}{2}\frac{n^{2}}{M_{A}}f_{\mu \nu
}n^{\mu }\partial ^{\nu }a^{2}  \label{lag0}
\end{equation}%
$\delta $ is Lagrange multiplier setting orthogonality condition for the
vector Goldstone field, thus treating it as gauge fixing one. In general, we
have here pletora Lorentz and CPT violating couplings like $\frac{n^{2}}{%
M_{A}}f_{\mu \nu }n^{\mu }\partial ^{\nu }a^{2}$ in the higher orders,
especially if charged currents are introduced as well, but it appears in all
physical processes (photon-photon, matter-photon, matter-matter
interactions) at least in the tree and one loop level, there is no sign of
physical Lorentz invariance violation. Looks like Lorentz invariance is
realized in nonlinear fashion and Lorentz breaking condition (\ref{c1}) is
treated like a nonlinear gauge choice for vector field \cite{az,kep}.

Consideration of the spontaneous Lorentz violation scenarios for non-Abelian
vector fields meet same challenges, though consequently lead to the same
conclusions as in the Abelian vector field case, despite the fact that there
are some significant differences as well. The length fixing constraint
adapted for non-Abelian vector fields in fact violates not only Lorentz
symmetry, but an accidental symmetry $SO(N,3N)$ of the constraint itself
(here $N$ defines unitary symmetry group of vector fields) which is much
higher than symmetry of the theory Lagrangian. This gives extra massless
modes which together with the true Lorentzian Goldstone complete the whole
gauge multiplet of the non-Abelian theory taken \cite{jej}.

\subsection{Emergent tensor field gravity}

Actually, for the tensor field gravity we can use the similar nonlinear
constraint for a symmetric two-index tensor field
\begin{equation}
H_{\mu \nu }^{2}=\mathfrak{n}^{2}M_{H}^{2}\text{ , \ }H_{\mu \nu }^{2}\equiv
H_{\mu \nu }H^{\mu \nu }\text{, \ }\mathfrak{n}^{2}\equiv \mathfrak{n}_{\mu
\nu }\mathfrak{n}^{\mu \nu }=\pm 1\text{ }  \label{const3}
\end{equation}%
(where $\mathfrak{n}_{\mu \nu }$ is now a properly oriented unit Lorentz
tensor, which supposedly specifies vacuum expectation values, while $M_{H}$
is the proposed scale for Lorentz violation in the gravity sector) which
fixes its length in the same manner as it appears for the vector field (\ref%
{c1}). Again, the nonlinear constraint (\ref{const3}) may in principle
appear from the standard potential terms added to the tensor field Lagrangian%
\begin{equation}
U(H)=\lambda _{H}(H_{\mu \nu }^{2}-\mathfrak{n}^{2}M_{H}^{2})^{2}\text{ }
\label{u11}
\end{equation}%
in the nonlinear $\sigma $-model type limit when the coupling constant $%
\lambda _{H}$ goes to infinity. Just in this limit the tensor field theory
appears stable, but doing so, we are effectively excluding corresponding
Higgs mode from the theory and it does not lead to physical Lorentz
violation \cite{cjt}.

This constraint (\ref{const3}), like the non-Abelian vector field, has
higher symmetry then the kinetic term, particularly $SO(7,3)$. So,
spontaneous symmetry violation breaks not only Lorentz symmetry, but also
this $SO(7,3)$ and therefore produces also PGM-s, but in contrast to vector
field, when we had only two channels of Lorentz symmetry violation to $SO(3)$
or $SO(1,2)$ and three true Goldstone modes always, for tensor field we have
more possibilities. If we write down constraint in more details
\begin{equation}
H_{\mu \nu }^{2}=H_{00}^{2}+H_{i=j}^{2}+(\sqrt{2}H_{i\neq j})^{2}-(\sqrt{2}%
H_{0i})^{2}=\mathfrak{n}^{2}M_{H}^{2}=\pm M_{H}^{2}
\end{equation}%
we see that if only one component of the tensor field should acquire vacuum
expectation value (assuming minimal vacuum configuration) we have following
alternatives: \
\begin{eqnarray}
(a)\text{ \ \ \ \ }\mathfrak{n}_{00} &\neq &0\text{ , \ \ }%
SO(1,3)\rightarrow SO(3)  \notag \\
(b)\text{ \ \ \ }\mathfrak{n}_{i=j} &\neq &0\text{ , \ \ }SO(1,3)\rightarrow
SO(1,2) \\
(c)\text{ \ \ \ }\mathfrak{n}_{i\neq j} &\neq &0\text{ , \ \ }%
SO(1,3)\rightarrow SO(1,1)  \notag
\end{eqnarray}

for $\mathfrak{n}^{2}=1$ and
\begin{equation}
(d)\text{ \ \ }\mathfrak{n}_{0i}\neq 0\text{ , \ \ }SO(1,3)\rightarrow SO(2)
\end{equation}%
for $\mathfrak{n}^{2}=-1$. For $a,$ $b$ cases we have three true Goldstone
modes and for $c$, $d$ we have five, since only one generator of Lorentz
transformations remains unbroken. While in $b,$ $c,$ $d$ cases physical
graviton consists, at least partially, from true Goldstone modes, in case $a$
only true goldstones are $H_{0i}$ components, thus physical graviton will be
constructed from PGM-s. One should notice that pseudo-Goldstone nature of
some components of tensor multiplet poses no threats and generally in
contrast to the scalar pseudo-Goldstone modes they do not acquire mass duo
to the quantum effects, if diffeomorphism (diff) invariance is present.

So, we are putting (\ref{const3}) on the tensor field mimicking linearized
general relativity
\begin{equation}
\mathrm{L}=L(H)+L_{S}-\frac{1}{M_{P}}H_{\mu \nu }T_{S}^{\mu \nu }  \label{tl}
\end{equation}%
where

\begin{equation}
L(H)=\frac{1}{2}\partial _{\lambda }H^{\mu \nu }\partial ^{\lambda }H_{\mu
\nu }-\frac{1}{2}\partial _{\lambda }H_{tr}\partial ^{\lambda
}H_{tr}-\partial _{\lambda }H^{\lambda \nu }\partial ^{\mu }H_{\mu \nu
}+\partial ^{\nu }H_{tr}\partial ^{\mu }H_{\mu \nu }\text{ }  \label{fp}
\end{equation}
Here $H_{tr}$ stands for the trace of $H_{\mu \nu }$ ($H_{tr}=\eta ^{\mu \nu
}H_{\mu \nu }$) and $L(H)$ is invariant under the diff transformations

\begin{equation}
\delta H_{\mu \nu }=\partial _{\mu }\xi _{\nu }+\partial _{\nu }\xi _{\mu }%
\text{ , \ \ \ }\delta x^{\mu }=\xi ^{\mu }(x)\text{ ,}  \label{tr3}
\end{equation}%
while $L_{S}$ and$\ T_{S}^{\mu \nu }$ are the Lagrangian and corresponding
energy momentum tensor of whatever is gravitating, (vector fields, matter).
In case, vector field is considered
\begin{equation}
L(A)=-\frac{1}{4}F_{\mu \nu }F^{\mu \nu },\text{ }T^{\mu \nu }(A)=-F^{\mu
\rho }F_{\rho }^{\nu }+\frac{1}{4}\eta ^{\mu \nu }F_{\alpha \beta }F^{\alpha
\beta }
\end{equation}%
where $L(H)$ is fully diff invariant, but that is not the case for other
parts of Lagrangian and diff invariance is satisfied only proximately, but
they become more and more invariant when the tensor field gravity Lagrangian
(\ref{tl}) is properly extended to GR with higher terms in $H$-fields
included\footnote{%
Such an extension means that in all terms included in the GR action,
particularly in the\ QED Lagrangian term , $(-g)^{1/2}g_{\mu \nu }g_{\lambda
\rho }F^{\mu \lambda }F^{\nu \rho }$, one expands the metric tensors
\begin{equation*}
g_{\mu \nu }=\eta _{\mu \nu }+H_{\mu \nu }/M_{P}\text{, \ }g^{\mu \nu }=\eta
^{\mu \nu }-H^{\mu \nu }/M_{P}+H^{\mu \lambda }H_{\lambda }^{\nu
}/M_{P}^{2}+\cdot \cdot \cdot
\end{equation*}%
taking into account the higher terms in $H$-fields.}.

Once tensor field acquires vacuum expectation value, we can expand it into
Goldstone mode
\begin{equation}
H_{\mu \nu }=h_{\mu \nu }+\mathfrak{n}_{\mu \nu }M_{H}-\frac{\mathfrak{n}%
^{2}h^{2}}{2M_{H}}+O(1/M_{H}^{2}),\text{ }\mathfrak{n}\cdot h=0\text{\ }
\label{par}
\end{equation}%
Here $h_{\mu \nu }$ corresponds to the pure emergent modes satisfying the
orthogonality condition and $h^{2}\equiv h_{\mu \nu }h^{\mu \nu }$, \ $%
\mathfrak{n}\cdot h\equiv \mathfrak{n}_{\mu \nu }h^{\mu \nu }$. \

Lets specify once again that $h_{\mu \nu }$ consists of Goldstone and PGM-s.
Only case, when physical graviton will consists of only Goldstone mode is
when Lorentz invariance is fully broken, we have six emergent goldstone
modes and other pseudo Goldstone components is gauged away by fixing
remaining gauge freedom (more about supplementary conditions below). Such a
scenario can not be achieved by minimal vacuum configuration. Nevertheless,
whether tensor field will be defined only by Goldstone modes or by a mixture
with PGM-s, hole tensor multiplet always stays strictly massless. A
particular case of interest is that of the traceless VEV tensor $\mathfrak{n}%
_{\mu \nu }$

\begin{equation}
\text{\ \ }\mathfrak{n}_{\mu \nu }\eta ^{\mu \nu }=0  \label{tll}
\end{equation}%
in terms of which the emergent gravity Lagrangian acquires an especially
simple form (see below). It is clear that the VEV in this case can be
developed on several $H_{\mu \nu }$ components simultaneously, which in
general may lead to total Lorentz violation with all six Goldstone modes
generated. For simplicity, we will use sometimes this form of vacuum
configuration in what follows, while our arguments can be applied to any
type of VEV tensor $\mathfrak{n}_{\mu \nu }$. \ \

Alongside to basic emergent orthogonality condition in (\ref{par}) one must
also specify other supplementary conditions for the tensor field $h^{\mu \nu
}$(appearing eventually as possible gauge fixing terms in the emergent
tensor field gravity). We have remaining three degrees of gauge freedom.
Usually, spin $1$ states in tensor field is gauged away by the conventional
Hilbert-Lorentz condition

\begin{equation}
\partial ^{\mu }h_{\mu \nu }+q\partial _{\nu }h_{tr}=0  \label{HL}
\end{equation}%
($q$ is an arbitrary constant, giving for $q=-1/2$ the standard harmonic
gauge condition), because spin-$1$ component always has negative
contribution in energy and therefore it is desirable action. However, as we
have already imposed the emergent constraint (\ref{par}), we can not use the
full Hilbert-Lorentz condition (\ref{HL}) eliminating four more degrees of
freedom in $h_{\mu \nu }.$ Otherwise, we would have an "over-gauged" theory
with a non-propagating graviton. In fact, the simplest set of conditions
which conform with the emergent condition $\mathfrak{n}\cdot h=0$ in (\ref%
{par}) turns out to be

\begin{equation}
\partial ^{\rho }(\partial _{\mu }h_{\nu \rho }-\partial _{\nu }h_{\mu \rho
})=0  \label{gauge}
\end{equation}%
This set excludes only three degrees of freedom \footnote{%
The solution for a gauge function $\xi _{\mu }(x)$ satisfying the condition$%
\ $(\ref{gauge}) $\ $\ can generally be chosen as $\xi _{\mu }=$\ $\ \square
^{-1}(\partial ^{\rho }h_{\mu \rho })+\partial _{\mu }\theta $, where $%
\theta (x)$ is an arbitrary scalar function, so that only three degrees of
freedom in $h_{\mu \nu }$ are actually eliminated.} in $h_{\mu \nu }$ and,
besides, it automatically satisfies the Hilbert-Lorentz spin condition as
well. \

Putting parameterization (\ref{par}) into the total Lagrangian given in (\ref%
{tl}), one comes to the truly emergent tensor field gravity Lagrangian
containing an infinite series in powers of the $h_{\mu \nu }$ modes. For the
traceless VEV tensor $\mathfrak{n}_{\mu \nu }$, without loss of generality,
we get the especially simple form \ \ \ \

\begin{eqnarray}
\mathrm{L} &=&\frac{1}{2}\partial _{\lambda }h^{\mu \nu }\partial ^{\lambda
}h_{\mu \nu }-\frac{1}{2}\partial _{\lambda }h_{tr}\partial ^{\lambda
}h_{tr}-\partial _{\lambda }h^{\lambda \nu }\partial ^{\mu }h_{\mu \nu
}+\partial ^{\nu }h_{tr}\partial ^{\mu }h_{\mu \nu }+  \notag \\
&&-\frac{\mathfrak{n}^{2}}{M_{H}}h^{2}\mathfrak{n}^{\mu \lambda }\left[
\partial _{\lambda }\partial ^{\nu }h_{\mu \nu }-\frac{1}{2}\partial _{\mu
}\partial _{\lambda }h_{tr}\right] +\frac{\mathfrak{n}^{2}}{8M_{H}^{2}}%
\left( \eta ^{\mu \nu }-\frac{\mathfrak{n}^{\mu \lambda }\mathfrak{n}^{\nu
\lambda }}{\mathfrak{n}^{2}}\right) \partial _{\mu }h^{2}\partial _{\nu
}h^{2}  \notag \\
&&+L_{S}-\left( M_{H}\mathfrak{n}_{\mu \nu }+h_{\mu \nu }-\frac{h^{2}%
\mathfrak{n}_{\mu \nu }}{2M_{H}}\right) \frac{T_{S}^{\mu \nu }}{M_{P}}%
+O(1/M_{H}^{2})  \label{lag1}
\end{eqnarray}

The bilinear field term \ \

\begin{equation}
\frac{M_{H}}{M_{P}}\mathfrak{n}_{\mu \nu }T_{S}^{\mu \nu }  \label{t}
\end{equation}%
in the third line in the Lagrangian (\ref{lag1}) merits special notice. This
term arises from the interaction term with tensor field. It could
significantly affect the dispersion relation for the all the fields included
in $T_{S}^{\mu \nu }$, thus leading to an unacceptably large Lorentz
violation if the SLIV scale $M_{H}$ were comparable with the Planck mass $%
M_{P}$. However, this term can be gauged away \cite{cjt} by an appropriate
redefinition of the fields involved by going to the new coordinates%
\begin{equation}
x^{\mu }\rightarrow x^{\mu }+\xi ^{\mu }.  \label{coo}
\end{equation}%
In fact, with a simple choice of the parameter function $\xi ^{\mu }(x)$
being linear in 4-coordinate

\begin{equation}
\xi ^{\mu }(x)=\frac{M_{H}}{M_{P}}\mathfrak{n}^{\mu \nu }x_{\nu }\text{ ,}
\label{df}
\end{equation}%
the term (\ref{t}) is cancelled by an analogous term stemming from the
kinetic term in $L_{S}$. On the other hand, since the diff invariance is an
approximate symmetry of the Lagrangian $\mathrm{L}$ we started with (\ref{tl}%
), this cancellation will only be accurate up to the linear order
corresponding to the tensor field theory. Indeed, a proper extension of this
theory to GR$^{1}$ with its exact diff invariance will ultimately restore
the usual dispersion relation for the vector field and other matter fields
involved. We will consider all that in significant detail in the next
section. \

So, with the Lagrangian (\ref{lag1}) and the supplementary conditions\ (\ref%
{par}) and (\ref{gauge}) lumped together, one eventually comes to a working
model for the emergent tensor field gravity \cite{cjt}. Generally, from ten
components of the symmetric two-index tensor $h_{\mu \nu }$ four components
are excluded by the supplementary conditions (\ref{par}) and (\ref{gauge}).
For a plane gravitational wave propagating in, say, the $z$ direction
another four components are also eliminated, due to the fact that the above
supplementary conditions still leave freedom in the choice of a coordinate
system, $x^{\mu }\rightarrow $ $x^{\mu }+\xi ^{\mu }(t-z/c),$ much as it
takes place in standard GR. Depending on the form of the VEV tensor $%
\mathfrak{n}_{\mu \nu }$, caused by SLIV, the two remaining transverse modes
of the physical graviton may consist solely of Lorentzian Goldstone modes or
of pseudo-Goldstone modes, or include both of them. This theory, similar to
the nonlinear QED \cite{nambu}, while suggesting an emergent description for
graviton, does not lead to physical Lorentz violation \cite{cjt}.

\subsection{Length Fixing Constraints and Nonlinear Gauge}

We have overviewed above the SLIV scenarios for vector and tensor fields and
could see that, though the well motivated length fixing constraint for a
given field causes spontaneous Lorentz violation, somewhat
counterintuitively, in physical processes, Lorentz symmetry appears intact.
Therefore we rightfully suspect that the Lorentz breaking constraint
condition acts effectively as a gauge fixing condition. To prove or disprove
whether this suspicion is reasonable one either should check the SLIV
effects in the corresponding physical processes in all orders, that looks
unrealistic, or has to find some generic argument, particularly find a
solution for gauge function or, at least, prove that such a solution exists.

In case of vector field $A_{\alpha }$ and Lorentz breaking condition $%
A_{\alpha }^{2}=n_{\beta }^{2}M_{A}^{2}$ , the corresponding equation for
gauge function $S$ is%
\begin{equation}
\left( A_{\alpha }+\partial _{\alpha }S\right) ^{2}=n_{\beta }^{2}M_{A}^{2}
\label{main.eq.V}
\end{equation}%
This equation is nonlinear and its exact solution for arbitrary $A_{\alpha }$
is not yet found. However, to our fortune, it is well known that this
equation taken for time-like violation case ($n_{\beta }^{2}=1$) is in fact
the Hamilton-Jacobi equation for the relativistic particle, which moves in
the external electromagnetic field. An action for such a system is given by
\begin{eqnarray}
S &=&\int M\sqrt{dx_{\alpha }dx^{\alpha }}-A_{\alpha }dx^{\alpha }  \notag \\
&=&\int \left( M\sqrt{u_{\alpha }u^{\alpha }}-A_{\alpha }u^{\alpha }\right)
d\tau
\end{eqnarray}%
where $\tau $ is evolution parameter and $u_{\alpha }=\dfrac{dx_{\alpha }}{%
d\tau }$. In this case, even though we do not have exact solution for that,
we know that an action $S$ describes a physical system and therefore it has
a solution for an arbitrary electromagnetic field field $A_{\alpha }$.

Analogously, for the space-like $n_{\beta }$ ($n_{\beta }^{2}=-1$) our basic
equation (\ref{main.eq.V}) might be considered as the Hamilton-Jacobi
equation for a hypothetical tachyon moving in the external electromagnetic
field
\begin{equation}
S=\int M\sqrt{-dx_{\alpha }dx^{\alpha }}-A_{\alpha }dx^{\alpha }=\int \left(
M\sqrt{-u_{\alpha }u^{\alpha }}-A_{\alpha }u^{\alpha }\right) d\tau
\end{equation}%
So, though this action can only correspond to a hypothetical particle, which
is not discovered so far, theoretically in might exist at least as a free
particle state. At this point we are unable to solve (\ref{main.eq.V})
exactly nor for time-like, neither for space-like cases, but we can check
that ultra-relativistic particle and tachyon (in the limit of very large
momenta, when particle velocity $v_{p}\longrightarrow c$ from below and
tachyon velocity\ $v_{t}\longrightarrow c$ from above) have \ somewhat
similar equations of motions
\begin{eqnarray}
\frac{d}{dt}p_{i} &=&F_{0i}-\frac{p_{_{l}}}{\sqrt{p_{_{k}}^{2}}}F_{li}
\notag \\
\frac{d}{dt}p_{i} &=&-F_{0i}+\frac{p_{_{l}}}{\sqrt{p_{_{k}}^{2}}}F_{li}
\end{eqnarray}%
with the electromagnetic field flipped for tachyon ($p_{i}$ stands for the
corresponding three-momenta). No dependent, one believes or not in an
existence of charged tachyon one might at least can take this similarity as
a hint that in space-like case, similar to a time-like violation, we are
dealing with effectively nonlinear gauge fixing condition.

For the tensor field, diff gauge invariance also could only fully be
approved, when corresponding gauge function $\xi _{\alpha }(x_{\mu })$ is
found, which\ satisfies the following equation
\begin{equation}
\left( H_{\alpha \beta }+\partial _{\alpha }\xi _{\beta }+\partial _{\beta
}\xi _{\alpha }\right) ^{2}=\pm M_{H}^{2}  \label{tc}
\end{equation}%
While we do not have a heuristic argument like that we had above for the
vector field time-like SLIV case, we can provide some arguments very similar
to its space-like violation case leading again to the mainly intuitive
suggestion.

So, to conclude, though the above discussion looks highly suggestive towards
the vector and tensor field constraints, (\ref{c1}) and (\ref{const3}), to
consider them as the be nonlinear gauge choices, they are not yet, sure, the
rigorous proofes. Therefore, presently the only way to check whether these
constraints are just gauge choices or not is actually related to seeking of
the SLIV efects by a direct analysis of the corresponding physical processes.

\section{Electrogravity model}

Usually, an emergent gauge field framework is considered either regarding
emergent photons or regarding emergent gravitons, but in nature they do not
exist in separate framework, they are different parts of one picture and
therefore the most natural thing is to discuss them as such. For the first
time, we consider it regarding them both in the so-called electrogravity
theory where together with the Nambu QED model \cite{nambu} with its gauge
invariant Lagrangian we propose the linearized Einstein-Hilbert kinetic term
for the tensor field preserving a diff invariance (more details can be found
in our recent paper \cite{epc}). We show that such a combined SLIV pattern,
conditioned by the constraints (\ref{c1}) and (\ref{const3}), induces the
massless Goldstone modes which appear shared among photon and graviton. One
needs in common nine zero modes both for photon (three modes) and graviton
(six modes) to provide all necessary (physical and auxiliary) degrees of
freedom. They actually appear in our electrogravity theory due to
spontaneous breaking of high symmetries of our constraints. While for a
vector field case the symmetry of the constraint coincides with Lorentz
symmetry $SO(1,3)$, the tensor field constraint itself possesses much higher
global symmetry $SO(7,3)$, whose spontaneous violation provides a sufficient
number of zero modes collected in a graviton. As we understand already these
modes are largely pseudo-Goldstone modes since $SO(7,3)$ is symmetry of the
constraint (\ref{const3}) rather than the electrogravity Lagrangian whose
symmetry is only given by Lorentz invariance.

\subsection{Constraints and zero mode spectrum}

Before going any further, let us make some necessary comments. Note first of
all that, apart from dynamics that will be described by the total
Lagrangian, the vector and tensor field constraints (\ref{c1}, \ref{const3})
are also proposed to be satisfied. In principle, these constraints, like in
previous cases, could be formally obtained from the conventional potential
introduced in the total Lagrangian. The most general potential, where the
vector and tensor field couplings possess the Lorentz and $SO(7,3)$
symmetry, respectively, must be solely a function of $A_{\mu }^{2}\equiv
A_{\mu }A^{\mu }$\ and $H_{\mu \nu }^{2}\equiv H_{\mu \nu }H^{\mu \nu }$.
Indeed, it cannot include any contracted and intersecting terms like as $%
H_{tr}$, \ $H^{\mu \nu }A_{\mu }A_{\nu }$ and others which would immediately
reduce the above symmetries to the common Lorentz one. So, one may only
write
\begin{equation}
U(A,H)=\lambda _{A}(A_{\mu }^{2}-n^{2}M_{A}^{2})^{2}+\lambda _{H}(H_{\mu \nu
}^{2}-\mathfrak{n}^{2}M_{H}^{2})^{2}+\lambda _{AH}A_{\mu }^{2}H_{\rho \nu
}^{2}  \label{u}
\end{equation}%
where $\lambda _{A,H,AH}$ stand for the coupling constants of the vector and
tensor fields, while values of $n^{2}=\pm 1$ and $\mathfrak{n}^{2}=\pm 1$
determine their possible vacuum configurations. As a consequence, an
absolute minimum of the potential (\ref{u}) might appear for the couplings
satisfying the conditions%
\begin{equation}
\lambda _{A,H}>0\text{ , \ }\lambda _{A}\lambda _{H}>\lambda _{AH}/4
\end{equation}%
However, as in the pure vector field case discussed in section 1, this
theory is generally unstable with the Hamiltonian being unbounded from below
unless the phase space is constrained just by the above nonlinear conditions
(\ref{c1}, \ref{const3}). They in turn follow from the potential (\ref{u})
when going to the nonlinear $\sigma $-model type limit $\lambda
_{A,H}\rightarrow \infty $. In this limit, the massive Higgs mode disappears
from the theory, the Hamiltonian becomes positive, and one comes to the pure
emergent electrogravity theory considered here.

We note again that the Goldstone modes appearing in the theory are caused by
breaking of global symmetries related to the constraints (\ref{c1}, \ref%
{const3}) rather than directly to Lorentz violation. Meanwhile, for the
vector field case symmetry of the constraint (\ref{c1}) coincides in fact
with Lorentz symmetry whose breaking causes the Goldstone modes depending on
the vacuum orientation vector $n_{\mu }$, as can be clearly seen from an
appropriate exponential parametrization for the starting vector field (\ref%
{2}). However, in the tensor field case, due to the higher symmetry $SO(7,3)$%
\ of the constraint (\ref{const3}), there are much more tensor zero modes
than would appear from SLIV itself. In fact, they complete the whole tensor
multiplet $h_{\mu \nu }$ in the parametrization (\ref{par}). However, as was
discussed in the previous section, only a part of them are true Goldstone
modes, others are pseudo-Goldstone ones. In the minimal VEV configuration
case, when these VEVs are developed only on the single $A_{\mu }$ and $%
H_{\mu \nu }$ components, one has several possibilities determined by the
vacuum orientations $n_{\mu }$ and $\mathfrak{n}_{\mu \nu }$. There appear
the twelve zero modes in total, three from Lorentz violation itself and nine
from a violation of the $SO(7,3)$\ symmetry that is more than enough to have
the necessary three photon modes (two physical and one auxiliary ones) and
six graviton modes (two physical and four auxiliary ones). We could list
below all possible cases corresponding $n-\mathfrak{n}$ values, \ the
timelike-spacelike SLIV, when $n_{0}\neq 0$ and $\mathfrak{n}_{i=j}\neq 0,$
the spacelike-timelike (nonzero $n_{i}$ and $\mathfrak{n}_{00}$),
spacelike-spacelike diagonal (nonzero $n_{i}$ and $\mathfrak{n}_{i=j}$) and
spacelike-spacelike nondiagonal (nonzero $n_{i}$ and $\mathfrak{n}_{i\neq j}$%
) cases, but for brevity, instead we only list the most interesting cases
corresponding to minimal and maximal Lorantz symmetry breaking.

(1) When both $n_{\mu }\neq 0$ and $\mathfrak{n}_{\mu \mu }\neq 0$, wether $%
\mu $ is time or space component we have minimally broken Lorentz invariance
and only three broken generators and therefore three Goldstone\ modes and
all of them is collected into the photon, while components of $h_{\alpha
\beta }$ needed for physical graviton and its auxiliary components can be
only provided by the pseudo-Goldstone modes following from the symmetry
breaking $SO(7,3)$ $\rightarrow SO(6,3)$ related to the tensor-field
constraint (\ref{const3}).

(2) For the case, when $n_{i}\neq 0$ and $\mathfrak{n}_{\beta \gamma }\neq 0$
(one of the nondiagonal space components of the unit tensor $\mathfrak{n}%
_{\mu \nu }$ is nonzero), when $i\neq \beta \neq \gamma $ Lorentz symmetry
appears fully broken so that the photon $a_{\mu }$ has three Goldstone
components , while the graviton is collected by the rest of true Goldstone
and PGM--s.

(3) Only case when both physical photon and graviton $h_{ij}$ consists of
true Goldstone modes is when $n_{0}\neq 0$ and $\mathfrak{n}_{i\neq j}\neq 0
$, but some gauge degrees of freedom for a graviton are given by the PGM
states stemming from the symmetry breaking of the tensor-field constraint (%
\ref{const3}).

In any case, while photon may only contain true Goldstone modes, some PGM-s
appear necessary to be collected in graviton together with some true
Goldstone modes to form full tensor multiplet.

\subsection{The Model}

In the previous section and Generally in emergent tensor field gravity
theories we considered the vector field $A_{\mu }$ as an unconstrained
material field which the emergent gravitons interacted with, but now in
electrogravity model we propose that the vector field also develops the VEV
through the SLIV constraint (\ref{c1}), thus generating the massless vector
Goldstone modes\ associated with a photon. We also include the complex
scalar field $\varphi $ (taken to be massless, for simplicity) as an actual
matter in the theory%
\begin{equation}
\mathcal{L}(\varphi )=D_{\mu }\varphi \left( D_{\mu }\varphi \right) ^{\ast
},\text{ }D_{\mu }=\partial _{\mu }+ieA_{\mu }\text{ .}  \label{fi}
\end{equation}%
So, the proposed total starting electrogravity Lagrangian is
\begin{equation}
\mathcal{L}_{tot}=L(A)+L(H)+L(\varphi )+L_{int}(H,A,\varphi )  \label{tot}
\end{equation}%
where $L(A)$ and $L(H)$ are U(1) gauge invariant and diff invariant vector
and tensor field Lagrangians, while the gravity interaction part%
\begin{equation}
L_{int}(H,A,\varphi )=-\frac{1}{M_{P}}H_{\mu \nu }[T^{\mu \nu }(A)+T^{\mu
\nu }(\varphi )]  \label{in}
\end{equation}%
contains the tensor field couplings with canonical energy-momentum tensors
of vector and scalar fields.

In the symmetry broken phase one goes to the pure Goldstone vector and
tensor modes, $a_{\mu }$ and $h_{\mu \nu }$, respectively, Which is
thoroughly discussed in the previous sections (\ref{lag0}), (\ref{lag1}). At
the same time, the scalar field Lagrangian $\mathcal{L}(\varphi )$ in (\ref%
{tot}) is going now to
\begin{equation}
\mathcal{L}(\varphi )=\left\vert \left( \partial _{\mu }+iea_{\mu
}+ieM_{A}n_{\mu }-ie\frac{n^{2}}{2M_{A}}a^{2}n_{\mu }\right) \varphi
\right\vert ^{2}  \label{ls}
\end{equation}%
while tensor field interacting terms (\ref{in}) in $\mathcal{L}%
_{int}(H,A,\varphi )$ convert to%
\begin{equation}
\mathcal{L}_{int}=-\frac{1}{M_{P}}\left( h_{\mu \nu }+M_{H}\mathfrak{n}_{\mu
\nu }-\frac{\mathfrak{n}^{2}}{2M_{H}}h^{2}\mathfrak{n}_{\mu \nu }\right) %
\left[ T^{\mu \nu }\left( a_{\mu }-\frac{n^{2}}{2M_{A}}a^{2}n_{\mu }\right)
+T^{\mu \nu }(\varphi )\right]  \label{lin}
\end{equation}%
where the vector field energy-momentum tensor is now solely a function of
the Goldstone $a_{\mu }$ modes.

\subsection{Emergent electrogravity interactions}

To proceed further, one should eliminate, first of all, the large terms of
the false Lorentz violation being proportional to the SLIV scales $M_{A}$
and $M_{H}$ in the interaction Lagrangians (\ref{ls}) and (\ref{lin}).
Arranging the phase transformation for the scalar field in the following way
\begin{equation}
\varphi \rightarrow \varphi \exp [-ieM_{A}n_{\mu }x^{\mu }]  \label{ph}
\end{equation}%
one can simply cancel that large term in the scalar field Lagrangian (\ref%
{ls}), thus coming to
\begin{equation}
\mathcal{L}(\varphi )=\left\vert \left( D_{\mu }-ie\frac{n^{2}}{2M_{A}}%
a^{2}n_{\mu }\right) \varphi \right\vert ^{2}  \label{ls1}
\end{equation}%
where the covariant derivative $D_{\mu }$ is read from now on as $D_{\mu
}=\partial _{\mu }+iea_{\mu }$. Another unphysical set of terms (\ref{t})
appear from the gravity interaction Lagrangian $L_{int}$ (\ref{lin}) where
the large SLIV\ entity $M_{H}\mathfrak{n}_{\mu \nu }$ couples to the
energy-momentum tensor. They also can be eliminated by going to the new
coordinates (\ref{coo}), as was mentioned in the previous section.

For infinitesimal translations $\xi _{\mu }(x)$ the tensor field transforms
according to (\ref{tr3}), while scalar and vector fields transform as%
\begin{equation}
\delta \varphi =\xi _{\mu }\partial ^{\mu }\varphi ,\text{ }\delta a_{\mu
}=\xi _{\lambda }\partial ^{\lambda }a_{\mu }+\partial _{\mu }\xi _{\nu
}a^{\nu }\text{ ,}  \label{st}
\end{equation}%
respectively. One can see, therefore, that the scalar field transformation
has only the translation part, while the vector one has an extra term
related to its nontrivial Lorentz structure. For the constant unit vector $%
n_{\mu }$ this transformation looks as
\begin{equation}
\delta n_{\mu }=\partial _{\mu }\xi _{\nu }n^{\nu },  \label{nt}
\end{equation}%
having no the translation part. Using all that and also expecting that the
phase parameter $\xi _{\lambda }$ is in fact linear in coordinate $x_{\mu }$
(that allows to drop out its high-derivative terms), we can easily calculate
all scalar and vector field variations, such as
\begin{equation}
\delta \left( D_{\mu }\varphi \right) =\xi _{\lambda }\partial ^{\lambda
}(D_{\mu }\varphi )+\partial _{\mu }\xi _{\lambda }D^{\lambda }\varphi ,%
\text{ }\delta f_{\mu \nu }=\xi _{\lambda }\partial ^{\lambda }f_{\mu \nu
}+\partial _{\mu }\xi ^{\lambda }f_{\lambda \nu }+\partial _{\nu }\xi
^{\lambda }f_{\mu \lambda }  \label{var}
\end{equation}%
and others. This finally leads to the total variations of the above
Lagrangians. Whereas the pure tensor field Lagrangian $L(H)$ (\ref{fp})\ is
invariant under diff transformations, $\delta L(H)=0$, the interaction
Lagrangian $L_{int}$ in (\ref{tot}) is only approximately invariant being
compensated (in the lowest order in the transformation parameter $\xi _{\mu
} $) by kinetic terms of all the fields involved. However, this Lagrangian
becomes increasingly invariant once our theory is extending to GR$^{1}$.

In contrast, the vector and scalar field Lagrangians acquire some nontrivial
additions%
\begin{eqnarray}
\delta L(A) &=&\xi _{\lambda }\partial _{\lambda }L(A)  \notag \\
&&-\frac{1}{2}\left( \partial _{\mu }\xi _{\lambda }+\partial _{\lambda }\xi
_{\mu }\right) \left[ f^{\mu \nu }f_{\nu }^{\lambda }+\frac{n^{2}}{M_{A}}%
\left( f_{\nu }^{\lambda }\partial ^{\mu \nu }a^{2}+\frac{1}{2}f_{\rho \nu
}\partial ^{\rho \nu }\left( a^{\mu }a^{\lambda }\right) \right) \right]
\notag \\
\text{ }\delta L(\varphi ) &=&\xi _{\lambda }\partial _{\lambda }L(\varphi
)+\left( \partial _{\mu }\xi _{\nu }+\partial _{\nu }\xi _{\mu }\right) %
\left[ \left( \mathfrak{D}^{\mu }\varphi \right) ^{\ast }\mathfrak{D}^{\nu
}\varphi +\frac{a^{\mu }a^{\nu }n^{2}}{2M_{A}}n_{\lambda }J_{\lambda }\right]
\label{ads}
\end{eqnarray}%
where $J_{\mu }$ stands for the conventional vector field source current
\begin{equation}
J_{\mu }=ie[\varphi ^{\ast }D_{\mu }\varphi -\varphi \left( D_{\mu }\varphi
\right) ^{\ast }]  \label{j}
\end{equation}%
while $\mathfrak{D}_{\nu }\varphi $ is the SLIV extended covariant
derivative for the scalar field
\begin{equation}
\mathfrak{D}_{\nu }\varphi =D_{\nu }\varphi -ie\frac{n^{2}}{2M_{A}}%
a^{2}n_{\nu }\varphi \text{ \ \ }  \label{not}
\end{equation}%
The first terms in the variations (\ref{ads}) \ are unessential since they
simply show that these Lagrangians transform, as usual, like as scalar
densities under diff transformations.

Combining these variations with $L_{int}$ (\ref{lin}) in the total
Lagrangian (\ref{tot}) one finds after simple, though long, calculations
that the largest Lorentz violating terms in it
\begin{equation}
-\left( \frac{M_{H}}{M_{P}}\mathfrak{n}_{\mu \nu }-\frac{\partial _{\mu }\xi
_{\lambda }+\partial _{\lambda }\xi _{\mu }}{2}\right) \left[ -f^{\mu \nu
}f_{\nu }^{\lambda }-\frac{n^{2}}{M_{A}}f_{\lambda }^{\nu }\partial ^{\mu
\lambda }a^{2}+2\mathfrak{D}^{\nu }\varphi \left( \mathfrak{D}^{\mu }\varphi
\right) ^{\ast }\right]  \label{11}
\end{equation}%
will immediately cancel if the transformation parameter is chosen exactly as
is given in (\ref{df}) in the previous section. So, with this choice we
finally have for the modified interaction Lagrangian

\begin{equation}
\mathcal{L}_{int}^{\prime }(h,a,\varphi )=-\frac{1}{M_{P}}h_{\mu \nu }T^{\mu
\nu }(a,\varphi )+\frac{1}{M_{P}M_{A}}\mathcal{L}_{1}+\frac{1}{M_{P}M_{H}}%
\mathcal{L}_{2}+\frac{M_{H}}{M_{P}M_{A}}\mathcal{L}_{3}  \label{intmod}
\end{equation}%
where
\begin{eqnarray}
\mathcal{L}_{1} &=&n^{2}h_{\mu \nu }\left[ f_{\lambda }^{\nu }\partial ^{\mu
\lambda }a^{2}-n^{\mu }J^{\nu }+\eta ^{\mu \nu }\left( -\frac{1}{4}%
f_{\lambda \rho }\partial ^{\lambda \rho }a^{2}+n^{\lambda }J_{\lambda
}\right) \right]  \notag \\
\mathcal{L}_{2} &=&\frac{1}{2}\mathfrak{n}^{2}h^{2}\mathfrak{n}_{\mu \nu }%
\left[ -f^{\mu \lambda }f_{\lambda }^{\nu }+2D^{\nu }\varphi \left( D^{\mu
}\varphi \right) ^{\ast }\right]  \notag \\
\mathcal{L}_{3} &=&n^{2}\mathfrak{n}_{\mu \lambda }\left[ \frac{1}{2}f_{\rho
\nu }\partial ^{\rho \nu }\left( a^{\mu }a^{\lambda }\right) -(a^{\mu
}a^{\lambda })n^{\nu }J_{\nu }\right]  \label{3}
\end{eqnarray}%
Thereby, apart from a conventional gravity interaction part given by the
first term in (\ref{intmod}), there are Lorentz violating couplings in $%
\mathcal{L}_{1,2,3}$ being properly suppressed by corresponding mass scales.
Note that the coupling presented in $\mathcal{L}_{3}$ between the vector and
scalar fields is solely induced by the tensor field SLIV. Remarkably, this
coupling may be in principle of the order of a normal gravity coupling or
even stronger, if \ $M_{H}>M_{A}$. However, appropriately simplifying this
coupling (and using also a full derivative identity) one comes to%
\begin{equation}
\mathcal{L}_{3}\sim n^{2}\left( \mathfrak{n}_{\mu \lambda }a^{\mu
}a^{\lambda }\right) n^{\rho }\left[ \partial ^{\nu }f_{\nu \rho }-J_{\rho }%
\right]
\end{equation}%
that after applying of the vector \ field equation of motion turns it into
zero. We consider it in more detail in the next section where we calculate
some tree level processes.

\section{The lowest order SLIV processes}

The emergent vector field Lagrangian (\ref{lag0}) and emergent gravity
Lagrangian in (\ref{lag1}) taken separately present in fact highly nonlinear
theory which contains lots of Lorentz and $CPT$ violating couplings.
Nevertheless, as it was shown in \cite{cjt,az,kep} in the lowest order
calculations, they all are cancelled and do not manifest themselves in
physical processes. As we talked about earlier, this may mean that the
length-fixing constraints (\ref{c1},\ref{const3})\ put on the vector and
tensor fields appear as the gauge fixing conditions rather than a source of
an actual Lorentz violation. \ In the context of electrogravity model, which
contains both photon and graviton as the emergent gauge fields, this means
that only source of new physics can be (\ref{intmod}). Even if suspicion
that length fixing constraints are nonlinear gauge choices is true, for
Lorentz invariance to be realized anyway, $U(1)$ and diff gauge
transformations should commute in the symmetry broken phase and then we
could claim that $\mathcal{L}_{1}$ and $\mathcal{L}_{2}$ in (\ref{intmod})
will have no physical effects, but there is also (\ref{3}), which is
proportional to diff transformation parameter and strictly speaking it is
not zero Lagrangian. So, in this picture to be logically sound and
consistent we should check all interactions in the (\ref{intmod}) anyway.

For that one properly derive all necessary Feynman rules and then calculate
the basic lowest order processes, such as photon-graviton scattering and
their conversion, photon scattering on the matter scalar field and other,
that has been throughly carried out in our paper mentioned above \cite{epc}
where can be found all necessary details. These calculations explicitly
demonstrate that all the SLIV effects in these processes are strictly
cancelled. This appears due to an interrelation between the longitudinal
graviton and photon exchange diagrams and the corresponding contact
interaction diagrams. So, physical Lorentz invariance in all processes is
left intact. Apart, many other tree level Lorentz violating processes
related to gravitons and vector fields (interacting with each other and the
matter scalar field in the theory) may also appear in higher orders in the
basic SLIV parameters $1/M_{H}$ and $1/M_{A}$, by iteration of couplings
presented in our basic Lagrangians (\ref{lag1}, (\ref{intmod})) or from a
further expansions of the effective vector and tensor field Higgs modes (\ref%
{3.0}, \ref{par}) inserted into the starting total Lagrangian (\ref{tot}).
Again, their amplitudes appear to cancel each other, thus eliminating
physical Lorentz violation in the theory.

Most likely, the same conclusion could be expected for SLIV\ loop
contributions as well. Actually, as in the massless QED case considered
earlier \cite{az}, the corresponding one-loop matrix elements in our
emergent electrogravity theory could either vanish by themselves or amount
to the differences between pairs of similar integrals whose integration
variables are shifted relative to each other by some constants (being in
general arbitrary functions of the external four-momenta of the particles
involved) which, in the framework of dimensional regularization, could lead
to their total cancellation.

So, after all, it should not come as too much of a surprise that emergent
electrogravity theory considered here is likely to eventually possess
physical Lorentz invariance provided that the underlying gauge and diff
invariance in the theory remains unbroken.

\section{Conclusion}

We have combined emergent photon and graviton into one framework of
electrogravity. While photon emerges as true vector Goldstone mode from
SLIV, graviton at least partially consists of PGM-s as well, because
alongside of Lorentz symmetry much bigger global symmetry of (\ref{const3}) $%
SO(7,3)$ is broken as well. Configuration of true Goldstone and PGM-s inside
graviton solely depends on VEV-s of vector and tensor fields. So, in total
12 massless Goldstone modes are born to complete photon and graviton
multiplets with an orthogonality conditions $n^{\mu }a_{\mu }=0$, \ $\mathit{%
n}^{\mu \nu }h_{\mu \nu }=0$ in place. Emergent electrogravity theory is
nonlinear and in principal contains many Lorentz and CPT violating
interactions, when expressed in terms of Goldstone modes. Nonetheless, all
non-invariant effects disappear in all possible lowest order physical
processes, which means that Lorentz invariance is intact and hence Lorentz
invariance breaking conditions (\ref{c1}, \ref{const3}) act as a gauge
fixing for photon and graviton, instead of being actual source of physical
Lorentz violation in the theory. If this cancellation occurs in all orders
(i.e. (\ref{c1}, \ref{const3}) are truly nonlinear gauge fixing conditions),
then emergent electrogravity is physically indistinguishable from
conventional gauge theories and spontaneous Lorentz violation caused by the
vector and tensor field constraints (\ref{c1}, \ref{const3}) appear hidden
in gauge degrees of freedom, and only results in a noncovariant gauge choice
in an otherwise gauge invariant emergent electrogravity theory.

From this standpoint, the only way for physical Lorentz violation to take
place would be if the above gauge invariance were slightly broken by near
Planck scale physics, presumably by quantum gravity or some other high
dimensional theory. This is in fact a place where the emergent vector and
tensor field theories may drastically differ from conventional QED,
Yang-Mills and GR theories where gauge symmetry breaking could hardly induce
physical Lorentz violation. In contrast, in emergent electrogravity such
breaking could readily lead to many violation effects including deformed
dispersion relations for all matter fields involved. Another basic
distinction of emergent theories with non-exact gauge invariance is a
possible origin of a mass for graviton and other gauge fields (namely, for
the non-Abelian ones, see \cite{jej}), if they, in contrast to photon, are
partially composed from pseudo-Goldstone modes rather than from pure
Goldstone ones. Indeed, these PGM-s are no longer protected by gauge
invariance and may properly acquire tiny masses, which still do not
contradict experiment. This may lead to a massive gravity theory where the
graviton mass emerges dynamically, thus avoiding the notorious discontinuity
problem \cite{zvv}.

So, while emergent theories with an exact local invariance are physically
indistinguishable from conventional gauge theories, there are some principal
distinctions when this local symmetry is slightly broken which could
eventually allow us to differentiate between the two types of theory in an
observational way.

\section*{Acknowledgements}

We would like to thank Colin Froggatt, Archil Kobakhidze and Holger Nielsen
for useful discussions and comments. Z.K. wants to thank participants of the
21st Workshop \textquotedblright What Comes Beyond the Standard
Models?\textquotedblright\ (23-30 June, Bled, Slovenia) for interesting and
useful discussions, as well as the organizers for such a productive and
working environment. This work is partially supported by Georgian National
Science Foundation (grant No. YS-2016-81).


\end{document}